# High-Pressure Crystal Growth, Superconducting Properties, and Electronic Band Structure of $Nb_2P_5$

Xiaolei Liu[†,±,§,#], Zhenhai Yu[†,#,*], Qifeng Liang[‡,*], Chunyin Zhou[⊥], Hongyuan Wang[†], Jinggeng Zhao[|], Xia Wang[†,⊥], Na Yu[†,⊥], Zhiqiang Zou[†,⊥], and Yanfeng Guo[†,&,*]

[†]School of Physical Science and Technology, ShanghaiTech University, Shanghai 201210, China
[±]Shanghai Institute of Optics and Fine Mechanics, Chinese Academy of Sciences, Shanghai 201800, China
[§]University of Chinese Academy of Sciences, Beijing 100049, China
[‡]Department of Physics, University of Shaoxing, Shaoxing 312000, China
[⊥]Shanghai Synchrotron Radiation Facility, Shanghai Advanced Research Institute, Chinese Academy of Sciences, Shanghai 201204, China
[|]Department of Physics, Harbin Institute of Technology, Harbin 150080, China
[⊥]Analytical Instrumentation Center, School of Physical Science and Technology, ShanghaiTech University, Shanghai 201210, China
[&]CAS Center for Excellence in Superconducting Electronics (CENSE), Chinese Academy of Sciences, Shanghai 200050, P. R. China

ABSTRACT: Orthorhombic (space group: *Pnma*) $Nb_2P_5$ is a high-pressure phase that is quenchable to ambient pressure, which could viewed as the zigzag $P_\infty$ chain-inserted $NbP_2$. We report herein the high-pressure crystal growth of $Nb_2P_5$ and the discovery of its superconducting transition at $T_c \sim 2.6$ K. The electrical resistivity, magnetization, and specific heat capacity measurements on the high-quality crystal unveiled a conventional type-II weakly coupled *s*-wave nature of the superconductivity, with the upper critical field $H_{c2}(0) \sim 0.5$ T, the electron-phonon coupling strength $\lambda_{ep} \sim 0.5 - 0.8$, and the Ginzburg-Landau parameter $\kappa \sim 100$. The *ab initio* calculations on the electronic band structure unveiled nodal-line structures protected by different symmetries. The one caused by band inversion, for example, on the Γ-X and U-R paths of the Brillouin zone, likely could bring nontrivial topology and hence possible nontrivial surface state on the surface. The surface states on the (100), (010) and (110) surfaces were also calculated and discussed. The discovery of the phosphorus-rich $Nb_2P_5$ superconductor would be instructive for the design of more metal phosphides superconductors which might host unconventional superconductivity or potential technical applications.

## INTRODUCTION

As one of the most active chemical elements, phosphorus can exist as isolated anion as well as in a network of polyanions with covalent P–P bond that usually appears in various metal phosphides with different stoichiometries ranging from metal-rich phosphides, like $T_3P$ (T denotes metal), to polyphosphides such as $TP_4$ (T = Cr, Mn, and Fe).[1] Metal phosphides representing a large family of materials have drawn continuous research interest due to the earth-abundance as well as their exotic physical and chemical properties. For example, varied physical properties from metallic behaviour to diamagnetic semiconductor or insulator in metal phosphides were observed, which are tightly related to the bonding states between metals and phosphorus.[2] In particularly, the superconductivity has always been one of the research focuses on metal phosphides, which is expected to provide opportunities for discovering unconventional pairing mechanisms. However, superconductors with chemical compositions of $T_xP_y$ are the minorities among the various families of superconductors. Superconductivity in these binary metal phosphides usually has an intimate relation with the interaction between the *s* and *p* electrons, mediated by screened plasma oscillations of the *d* electrons. The density of states (DOS) at the Fermi surface (FS) is very high but the electron-phonon interaction is very weak for the *d* electrons, which therefore only results in relatively low critical superconducting temperature ($T_c$) of most of the metal phosphides superconductors. For example, superconductivity found in $Mo_3P$, $Mo_8P_5$, and $Mo_4P_3$ is only with the $T_c$ of 7 K, 5.8 K and 3 K, respectively.[3] Since P has a very high vapour pressure, metal phosphides are consequently not easy to be prepared by conventional methods, especially those phosphorus rich metal phosphides. The high-pressure technique is efficient in synthesizing such compounds, such as NbPS ($T_c$ = 12 K),[4] etc. $ZrRu_4P_2$ synthesized at ~ 4 GPa and 850 °C with a tetragonal $ZrFe_4Si_2$-type structure ($P4_2/mnm$) shows superconductivity at ~ 11 K.[5] The $Fe_2P$-type MoNiP and $Co_2P$-type MoRuP prepared at ~ 4 GPa and 1600 °C have the $T_c$ of about 15.5 K,[6,7] respectively. Some metal phosphides or phosphides are not superconducting at normal pressure, but they could become superconductors through chemical doping or applying external pressure. For example, Rh-doped RuP ($Ru_{0.55}Rh_{0.45}P$) becomes superconducting at $T_c$ = 3.7 K [8] and the black phosphorus shows superconductivity at $T_c$ up to 9.5 K when a pressure of 32 GPa is applied.[9] Pressure-induced superconductivity was also reported in MnP ($T_c$ ~ 1.0 K at 8.0 GPa) and MoP ($T_c$ = 2.5 K at 30.0 GPa) without alternation of crystallographic symmetry.[10-14] These results point out another way toward the discovery of superconductivity in metal phosphides.

On the other side, the superconductivity in metal phosphides has recently been tied with nontrivial topological band structure. For example, WP was reported as the first superconductor ($T_c$ ~ 0.8 K) among 5d transition metal pnictides with the MnP-type structure at ambient pressure.[15] It was also predicted to be a topological high symmetry line semimetal when the spin-orbit coupling (SOC) is considered.[16] Furthermore, in MoP which crystallizes in the WC-type structure, triply degenerate fermion was theoretically predicated and then experimentally verified.[17] The nontrivial topological band structure in MoP and superconductivity are expected to coexist at pressure above 30 GPa according to the density functional theory (DFT) calculations.[14] However, the correlation between the superconductivity in these metal phosphides and the nontrivial topological states remains inconclusive, which is still waiting for further studies to establish.

$Nb_2P_5$ was ever synthesized by the high-pressure technique in 1980,[18] but the electrical and magnetic properties were only measured to ~ 77 K, which shows paramagnetism and a metallic conduction behaviour. In this paper, by optimizing the high-pressure synthesis conditions, we succeeded in growing high-quality $Nb_2P_5$ crystals and found a superconducting transition at around 2.6 K. The superconducting properties were studied in details and the electronic band structure was also investigated in details by the *ab initio* calculations.

## EXPERIMENTAL SECTION
### Crystal Growth

$Nb_2P_5$ crystals were grown by reacting Nb and red phosphorus under high-pressure conditions in a Kawai-type multi-anvil high-pressure apparatus (see Fig. S1). The mixture of Nb and P powders in a mole ratio of about 1 : 4 was compressed into a cylinder and placed in a hexagonal boron nitride (BN) cell with a size of 3.5 mm in inner diameter and 3.2 mm in depth. The cell was packaged into a MgO octahedral pressure medium with octahedral edge-length (OEL) of 14 mm and was then placed at the cube formed by eight WC anvils with truncation edge-length (TEL) of 8 mm. The sample assembly for the preparation of $Nb_2P_5$ adopted the same one as that by K. D. Leinenweber *et al*.,[19] which was first pre-compressed under a pressure of 3.5 GPa and was then heated up to 1100 °C with 20 minutes with maintaining the pressure. The temperature was measured by type-C WRe (W5%Re/W26%Re) thermocouples. After a reaction for two hours at this pressure and temperature, the sample was slowly cooled down to 900 °C within three hours, followed by quenching to room temperature by turning off the electrical power supply without releasing the pressure. The pressure was finally gradually released at room temperature.

### Crystal Characterization

Crystal structure of $Nb_2P_5$ was carefully identified by using the single crystal X-ray diffractometer (SXRD) equipped with a Mo $K_\alpha$ radioactive source ($\lambda$ = 0.71073 Å) at 150 K. The chemical compositions and uniformity of stoichiometry were checked by the energy dispersive spectroscopy (EDS) by using many arbitrarily selected micro-crystals cracked from a bulk crystal (see Fig. S2). The results revealed good stoichiometry of the crystals with Nb : P = 1.99 : 5. Using Olex2,[20] the structure of $Nb_2P_5$ was refined with the SHELXL[21] refinement package using Least Square minimization. The sample used for electrical transport measurements is a dense polycrystal which is obtained by pressing several small single crystals together under high pressure at 6 GPa without heating, which has the size of about 100 $\mu$m. The electrical resistivity and specific heat capacity measurements were carried out in a Quantum Design Physical Properties Measurement System (PPMS). The resistivity was measured in a standard four-probe configuration by attaching Au wires with silver paste on the sample. The schematic picture for the set-up is presented in Fig. S3. The direct-current (dc) magnetic susceptibility and isothermal magnetization were measured in a Quantum Design Magnetic Properties Measurement System (MPMS-3).

### The *ab initio* calculations

The first principles calculations were performed by the Vienna *ab initio* simulation package (VASP)[22] using the projected augmented-wave (PAW) method.[23,24] The exchange-correlation functional introduced by Perdew, Burke, and Ernzerhof (PBE)[25] within generalized gradient approximation (GGA) is adopted in the calculations. A gamma-centered 6×8×6 k-point mesh within a Monkhorst-Pack scheme is adopted for the Brillouin zone integrations. The energy cutoff for the plane-wave basis is set as 520 eV and the forces are relaxed less than 0.01 eV/Å. The k-mesh for the DFT calculation is 6 × 8 × 6. The atoms are allowed to relax while the lattice parameters of the unit cells are fixed to the experimental values. The surface density of state is calculated by using a recursive Green's function method[26,27] from a tight-binding models constructed by using the Maximally Localized Wannier Functions (MLWF) method coded in WANNIER90.[28]

## RESULTS AND DISSCUSSION

The $Nb_2P_5$ crystal used for the single crystal X-ray diffraction examination is shown in Fig. 1(a). The diffraction patterns shown in Figs. 1(b)-(d) was completely indexed on the basis of an orthorhombic unit cell with the lattice parameters $a$ = 16.7266 Å, $b$ = 3.3453 Å, $c$ = 7.9093 Å in the space group *Pnma* (No. 62), which are nicely consistent with those reported previously.[18] The goodness-of-fit on $F^2$ is 1.198. The final $R_1$ is 0.0164 (I > 2σ(I)) and $wR_2$ is 0.0330 (all data). The complete refinement results are summarized in table S1. $Nb_2P_5$ crystallizes into the orthorhombic structure with space group *Pnma* as shown in Fig. 2(d). In one unit cell of this structure, there are two crystallographically inequivalent Nb atoms, i. e. Nb(1) and Nb(2), locating at Wyckoff position 4c and five crystallographically inequivalent P atoms, P(1), P(2), P(3), P(4) and P(5), occupying Wyckoff position 4c. Each Nb atom is bonded with eight P atoms forming a polyhedron as shown in right panel of Fig. 2(**c**). The two polyhedrons comprising two different Nb sites are coplanar and superimposed along the b-axis direction to form the structure (see Fig. 2(d)). The perfect reciprocal space reflections without any other miscellaneous points, seen in Figs. 1(b)-(d), also indicate the high-quality of the crystal.

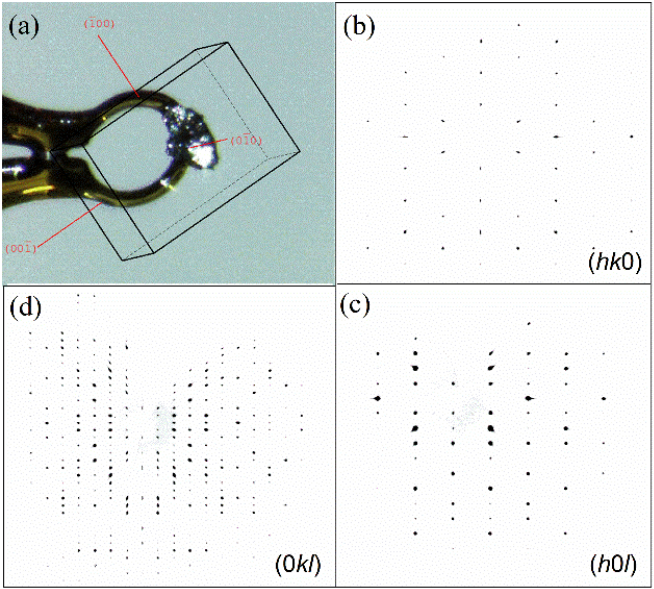

**Fig. 1.** (Color Online) (a) Optical image of $Nb_2P_5$ single crystal in single crystal X-ray diffraction measurement. (b)-(d) Diffraction patterns in the reciprocal space along the (h k 0), (h 0 l), and (0 k l) directions.

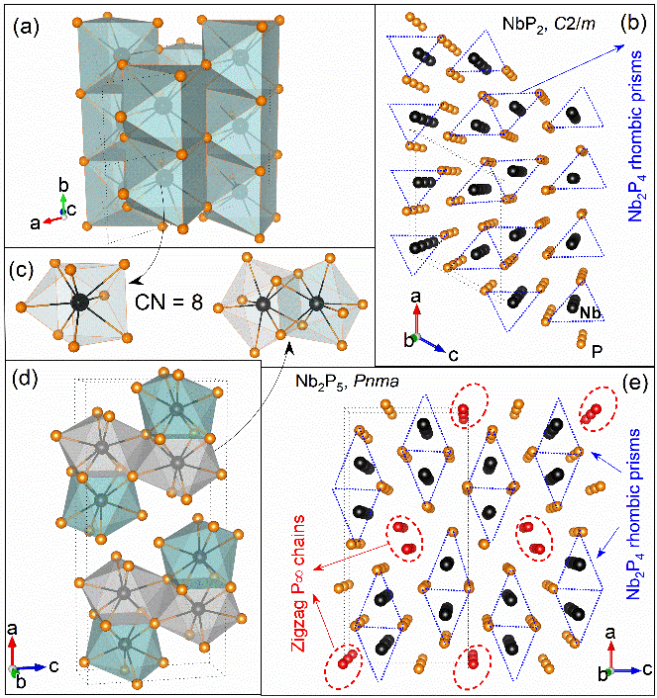

**Fig. 2.** (Color Online) Schematic crystal structure relationship between $NbP_2$ and $Nb_2P_5$. (a) and (b) The crystal structure of $NbP_2$ can be seen as composed with infinite rhombic prisms with the composition $Nb_2P_4$ stacking along b-axis. (c) The comparison of coordination polyhedron in $NbP_2$ and $Nb_2P_5$, and the coordination numbers (CNs) for Nb atoms in both of $NbP_2$ and $Nb_2P_5$ are 8. (d) and (e) The crystal structure of $Nb_2P_5$ is closely related to that of $NbP_2$, in which each zigzag $P_\infty$ chains are surrounded by four $Nb_2P_4$ rhombic prisms.

The crystal structure plays a critical role in determining the physical properties, such as magnetism, electronic band structure, and electrical conductivity (such as superconductivity). Therefore, it is valuable to pay a close attention onto the crystal structure of $Nb_2P_5$, which may provide useful insights into its superconducting properties. It is well known that $NbP_2$ is an intensively studied carrier compensated semimetal and one of the Nb-P compounds with the highest P content. $NbP_2$ crystallizes into the $OsGe_2$-type structure with a space group of C2/m (Fig. 2(a)). In $NbP_2$, seen in Fig. 2(a) and (c), each Nb atom is surrounded by eight P atoms and six such unit form a trigonal prism with two additional P atoms outside the rectangular faces. Two trigonal prisms form the $Nb_2P_4$ rhombic prisms parallelepiped, shown by the blue dashed triangles in Fig. 2(b), by sharing one of the three rectangular faces. These $Nb_2P_4$ rhombic prisms are stacked along the b-axis forming infinite layers. It can be seen from Fig. 2(e) that the crystal structure of $Nb_2P_5$ is closely related to the $NbP_2$, in which the infinite $Nb_2P_4$ rhombic prisms stacked along b-axis. But in the structure of $Nb_2P_5$, the zigzag $P_\infty$ (red dashed ellipses in Fig. 2(e)) chain enters the space which is surrounded by four $Nb_2P_4$ rhombic prisms. The comparison of coordination polyhedron in $NbP_2$ and $Nb_2P_5$ is shown in Fig. 2(c). The CNs for Nb atoms in both of $NbP_2$ and $Nb_2P_5$ are 8. However, the $NbP_8$ polyhedra are stacked with triangular and quadrilateral faces in $NbP_2$ and $Nb_2P_5$, respectively. These differences make $NbP_2$ and $Nb_2P_5$ very different in their physical properties. Moreover, the Nb(1)-Nb(2) atomic distance of 2.986 Å in $Nb_2P_5$ indicates the formation of Nb-Nb bonding. This bonding distance is quite close to that in $NbAs_2$, 3.00 Å. Keeping these in mind, we can briefly summarize the electric conduction of the isostructural $OsGe_2$-type compounds. The $MoAs_2$ and $WAs_2$ are superconducting at 0.4 K and 0.9 K,[29-31] respectively, while $NbAs_2$ and $TaAs_2$ do not show superconductivity.[32,33] The absence of superconductivity in $NbAs_2$ may be due to the low DOS at the Fermi level.[34] While when the pressure is applied, $NbAs_2$ can become a superconductor with a $T_c$ of 2.63 K under 12.8 GPa.[35] The main hole pocket of $NbAs_2$ encloses one time-reversal-invariant momenta in the electronic structure of the monoclinic lattice, suggesting $NbAs_2$ as a candidate for topological superconductor. The pressure study on $NbP_2$ has not been carried out yet. Since $Nb_2P_5$ could be viewed as the zigzag $P_\infty$ chain-inserted $NbP_2$, the superconductivity in $Nb_2P_5$ might have a relation with the extra $P_\infty$ chains.

Fig. 3 presents the magnetic susceptibility ($4\pi\chi$), resistivity ($\rho$), and specific heat capacity ($C_p$), which strongly confirm the bulk superconductivity in $Nb_2P_5$. The $\rho(T)$ measured at $B = 0$ T is shown by Fig. 3(a), which displays a gradual decrease upon cooling. Before the onset of superconducting transition, $\rho(T)$ shows a slight upturn with cooling, probably due to the disorder or impurities in $Nb_2P_5$ crystal that are commonly found in high-pressure synthesized crystals. The onset superconducting transition is observed at 2.7 K with a transition width less than 0.4 K with a relatively low residual resistivity ratio that may be attributed to electrons scattering effect. The details could be clearly seen by the enlarged view as an inset in Fig. 3(a). Seen in Fig. 3(b), the magnetic susceptibility measured in the temperature ($T$) range of 1.8 - 5 K under a magnetic field $B$ of 5 Oe exhibits diamagnetic signal below $T_c$ ~ 2.6 K, close to the $T_c$ determined from $\rho(T)$. Considering that the demagnetization factor $N$ equals to 0.286 due to the rectangular sample shape,[36] the largest shielding volume fraction in zero-field-cooling (ZFC) measurement is up to 82.3% at 1.8 K, implying bulk superconductivity in $Nb_2P_5$. The $C_p(T)$ at $B = 0$ further demonstrates the bulk superconductivity of $Nb_2P_5$ crystal, shown by the large jump at 2.6 K in Fig. 3(c), harmonically consistent with the values determined from both of $4\pi\chi$ and $\rho(T)$.

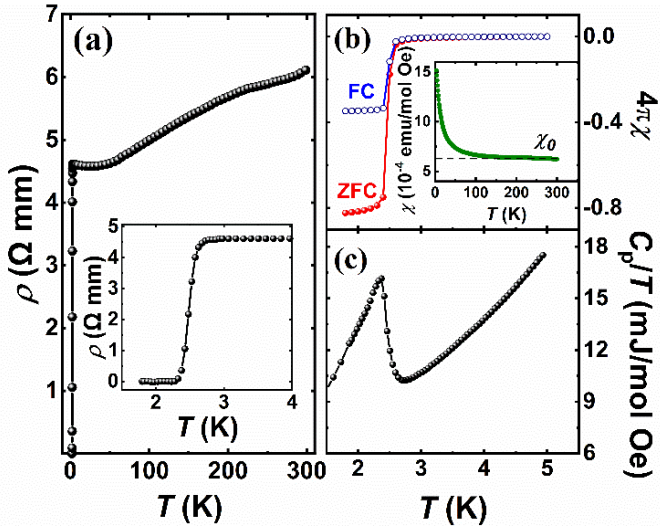

**Fig. 3.** (Color Online) (a) The temperature dependent resistivity $\rho(T)$ of $Nb_2P_5$ at 1.8 - 300 K. Inset shows an enlarged view near the superconducting transition. (b) The dc magnetic susceptibility of $Nb_2P_5$ measured in ZFC (closed circles) and FC mode (open circles) at a magnetic field of 5 Oe. Inset shows the data collected within 1.8 - 300 K at 40 kOe. (c) The $C_p/T$ near the $T_c$.

To achieve in-depth insights into the superconductivity of $Nb_2P_5$, the critical superconducting parameters were estimated by analyzing the $C_p(T)$. The $C_p(T)$ at $B = 0$ T within the temperature range of 3.5 - 5 K was fitted by using $C_p(T)/T = \gamma_n + \beta T^2$, where the first and second terms represent electronic and lattice contributions, respectively. As is shown in Fig. 4(a), the fitting gives $\gamma_n = 6.6674$ mJ mol$^{-1}$ K$^{-2}$, $\beta = 0.4457$ mJ mol$^{-1}$ K$^{-4}$. The Debye temperature $\Theta_D$ was estimated as 127 K through $\Theta_D = \sqrt[3]{\frac{12\pi^4 nR}{5\beta}}$ where $n = 7$ is the number of atoms per formula unit and $R$ is the molar gas constant. The electron-phonon coupling constant ($\lambda_{ep}$) can be obtained from the McMillan's relation[37]

$$\lambda_{ep} = \frac{1.04 + \mu^* \ln\left(\frac{\Theta_D}{1.45 T_C}\right)}{(1 - 0.62\mu^*)\ln\left(\frac{\Theta_D}{1.45 T_C}\right) - 1.04},$$

where $\mu^*$ is the Coulomb repulsion constant. Taking $\mu^*$ in the range of 0.05 - 0.2, $\lambda_{ep}$ is estimated to be 0.5 - 0.8, implying a weak coupling superconductivity in $Nb_2P_5$. The electronic specific heat capacity, denoted by $C_{el}(T)$, was obtained by subtracting the lattice contribution $\beta T^3$ from $C_p(T)$, which is presented in Fig. 4(b) as $C_{el}/T$ versus $T$. The $\Delta C_{el}/\gamma_n T_c$ is estimated to be 1.12, slightly lower than the value of 1.43 predicated by the BCS theory. The low temperature $C_{el}(T)$ to 0.5 K was fitted by the expression $C_{el}/T = \gamma_{res} + A\exp(-\beta\Delta)$,[38] where $\gamma_{res}$ is the residual Sommerfeld coefficient from the nonsuperconducting fraction of the sample and the second term represents a phenomenological exponential decay expected for an $s$-wave superconductor. The fitting result shown in Fig. 4(b) by the solid curve gives $\gamma_{res} = 1.51$ mJ mol$^{-1}$ K$^{-2}$, suggesting that the superconducting volume of the crystal is about 77%, which is close to that estimated from the magnetic susceptibility. The nice fit by the $s$-wave model suggests $Nb_2P_5$ as a conventional $s$-wave superconductor. Given both $\gamma_n$ (6.6674 mJ mol$^{-1}$ K$^{-2}$) and $\chi_0$ (6.1 × 10$^{-4}$ emu/mol Oe), the Wilson ratio can be estimated, giving $R = \chi_0/(3\gamma_n)(\pi k_B/\mu_B)^2 \approx$ 6.6, where $\chi_0$ is the temperature-independent susceptibility, $k_B$ is the Boltzmann's constant and $\mu_B$ is the Bohr magneton. The value is significantly larger than that expected for free electron value, $R = 1$, which could be attributed to strong electronic correlations or the other extrinsic effects.

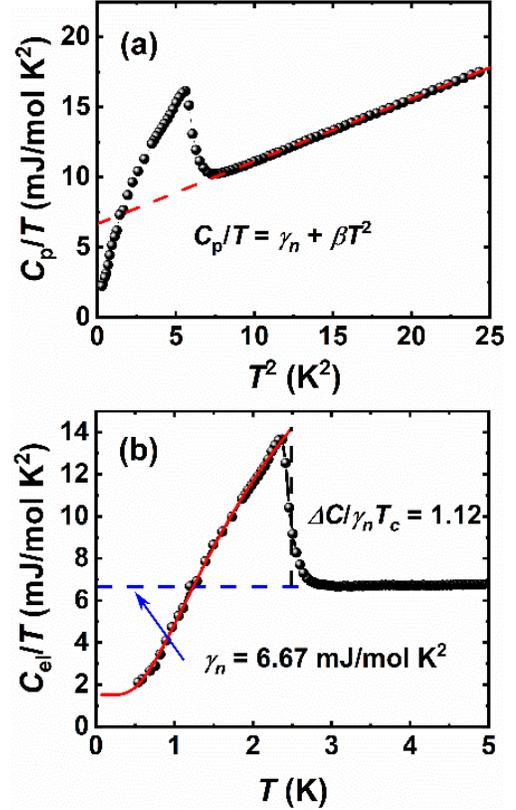

**Fig. 4.** (Color Online) (a) The $C_p(T)/T$ vs. $T^2$ for $Nb_2P_5$. The red dash line shows the fit result by using $C_p(T)/T = \gamma_n + \beta T^2$. (b) The $C_{el}/T$ vs. $T$. The solid curve represents the fit by using the single band isotropic s-wave BCS model. The entropy balance between superconducting and normal state at $T_c$ is maintained during the fit.

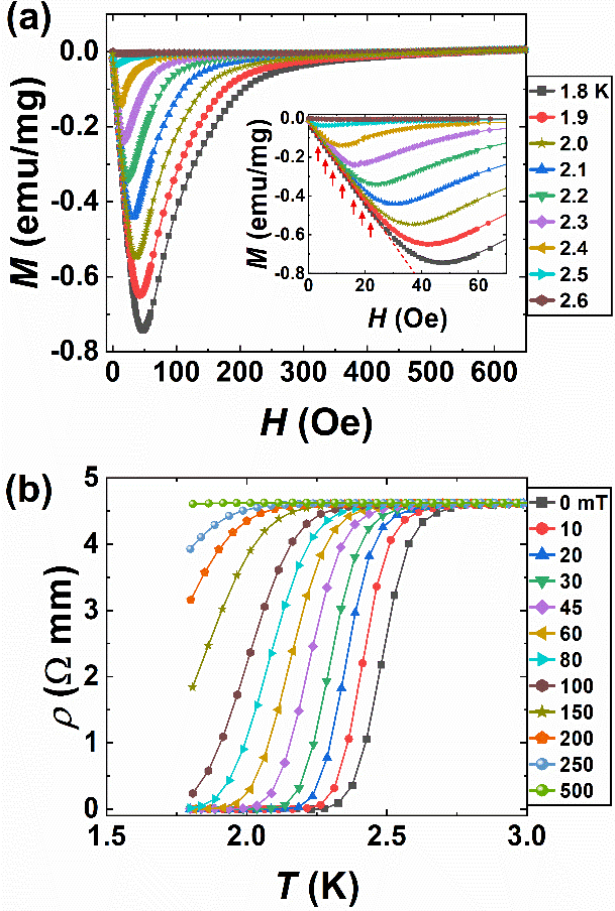

**Fig. 5.** (Color Online) (a) The isothermal magnetizations measured at $T = 1.8 - 2.6$ K with an interval of 0.1 K. Inset enlarges the low magnetic field region, with the position of $H_{c1}(T)$ being marked by red arrows. (b) $\rho(T)$ data under various applied magnetic fields of $H = 0, 10, 20, 30, 45, 60, 80, 100, 150, 200, 250$ and 500 Oe.

The field dependent magnetization and the low temperature resistivity of $Nb_2P_5$ are depicted in Fig. 5. The isothermal magnetizations measured at $T = 1.8 \sim 2.6$ K shown in Fig. 5(a) exhibit a typical type-II superconductor behavior by the clear hysteresis loops. The lower critical field $H_{c1}$ could be determined from the magnetization, which corresponds to the field values where the magnetization starts to depart from the linear evolution, as indicted by the arrows shown by the inset in Fig. 5(a). Fig. 5(b) shows the magnetic field dependence of $\rho$, which displays a clear shift of the $T_c$ to lower temperature with increasing the magnetic field. By locating the points where $\rho$ drops to onset or middle of its normal state value $\rho_n$ just above $T_c$, we can construct the $H_{c2}$ - $T$ phase diagram, shown in Fig. 6. Based on the Ginzburg-Landau (GL) equation[39]

$$H_{c2}(T) = H_{c2}(0)\frac{1-t^2}{1+t^2},$$

where $t = T/T_c$ is the reduced temperature. The fitting result is shown by the red solid line in Fig. 6. Werthamer-Helfand-Hohenberg (WHH) model was also applied to analyze the upper critical field, shown by the black dashed line.[40] In the $T_c^{onset}$ criteria, the zero-temperature upper critical field $H_{c2}(0)$ determined by GL and WHH models are 0.7 T and 0.59 T, respectively. In the $T_c^{mid}$ criteria, the two values of $H_{c2}(0)$ drop to 0.49 T and 0.37 T, respectively. Using the above critical field values to estimate the GL parameter $\kappa = \lambda/\xi \approx H_{c2}/H_{c1}$ (where $\lambda$ is the penetration depth, and $\xi$ represents the coherence length), it can be concluded that $Nb_2P_5$ is in the extreme type-II limit, as $\kappa \approx 0.2$ T/2 mT = 100.[41]

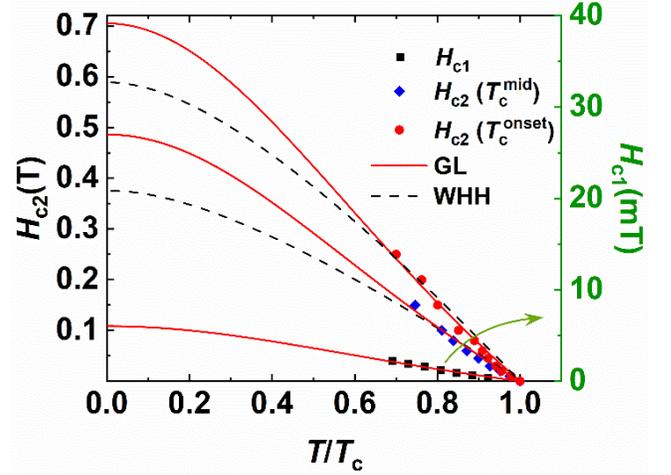

Fig. 6. (Color Online) Upper (lower) critical field $H_{c2}$ ($H_{c1}$) versus the normalized temperature $T/T_c$. $H_{c2}$ determined from onset and middle of transition around $T_c$ are all presented. The fitted curves using the WHH (black dashed line) and GL (red solid line) models are also displayed.

Study on the electronic band structure of $Nb_2P_5$ would be helpful for understanding the superconducting properties. The *ab initio* band structures along the high-symmetry paths of the bulk Brillouin Zone (BZ) are plotted in Fig. 7 (a), which indicate that the material is a metal. Seen in Fig. 7(a), around the Fermi level, the bands are mainly denoted by the $d_{xz}$, $d_{yz}$ and $d_{x^2-y^2}$ orbits of Nb, though their hybridization with the $p$ orbits of P could not be negligible.

As we mentioned above, $Nb_2P_5$ takes a space group of *Pnma* for its crystalline structure, which contains the space inversion P, three off-center mirror symmetries $G_z$:{$M_z|\frac{1}{2} 0 \frac{1}{2}$}, $M_{y'}$:{$M_y|0 \frac{1}{2} 0$} and $G_x$:{$M_x|0 \frac{1}{2} \frac{1}{2}$}, and three skew axes $S_{2z}\{C_{2z}|\frac{1}{2} 0 \frac{1}{2}\}$, $S_{2y}\{C_{2y}|0\frac{1}{2}0\}$ and $S_{2x}\{C_{2x}|\frac{1}{2}\frac{1}{2}\frac{1}{2}\}$.

In the absence of SOC, the combination of these symmetries enforces a two-fold degeneracy in the whole BZ, four-fold degeneracies at high symmetry paths SX, SY, UZ, UX, UR, TZ, TR and eight-fold degeneracies at SR and TY. These symmetries also induce band crossings around the Fermi energy, which have been highlighted by dotted circles in Fig. 7(a). The details of band crossing are also shown by Figs. 7 (c)-(f). By examining the band touch points of two nearest bands on the Fermi surface, the formed nodal-line structure is shown in Fig. 7(a). By inspecting the location of the nodal-lines, one can figure out the protection mechanism. For example, the band crossing on path Γ-X sits on a nodal-line that circles around the Γ point in the plane $k_z = 0$, implying a protection given by glide-reflection $G_z$. The band crossings on the U-R path, on the other hand, locate on a nodal-line which does not lie on any high-symmetry path or plane and the protection is from the *PT* (*T* denoted the time reversal symmetry) symmetry. From the band structure of Nb$_2$P$_5$ presented in Fig. 7(a), the band crossing on the Γ-X and U-R paths will be eliminated via tuning the energies of the two crossing bands. Such band-inversion nature might bring nontrivial topology for those crossing bands and possibly manifests nontrivial surface state on the surface of a slab system. However, it remains a possibility before a further careful check with symmetry analysis, which will not be a point that should be emphasized in this work. When the SOC is included, the nodal-lines protected by the *PT* symmetry and mirror symmetry will be gapped out. It has been shown in the Figs. 7 (c) and (e) for the gap-opening of the band crossings for instance.

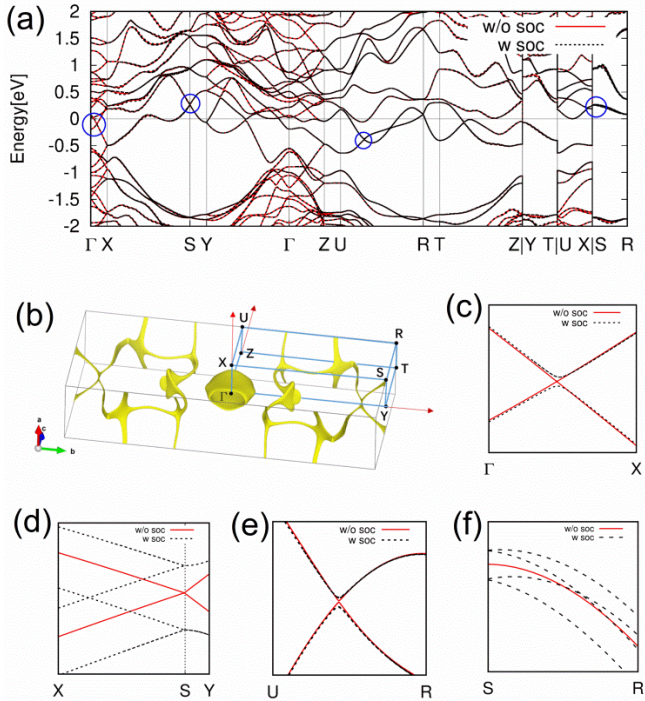

**Fig. 7.** (Color Online) (a) The energy bands of Nb$_2$P$_5$ with (dashed line) and without SOC (solid line). (b) The first Brillouin zone and the band crossing of two energy bands on the Fermi surface. The typical crossing points are highlighted by dashed circles (blue) and their details are shown in (c)-(f).

Because of its non-symmorphic symmetries, Nb$_2$P$_5$ also exhibits another type of nodal-line which is stable against the SOC interaction. In the Figs. 7(d) and (f) we plot the detailed band structure around S. Inclusion of SOC splits the eight-fold degeneracy of S into two four-fold degenerate states. Around the S point, the emerged new band crossings form a tiny nodal-line circles around S on the plane $k_b = \pi$. Such nodal-line is enforced by the glide reflection $G_b$ and the skew axes $S_z$, which was already reported previously in another *Pnma* crystal SrIrO$_3$.[42] Because of the relatively light masses of P and Nb, inclusion of SOC only alternated the band structure slightly (see in Fig. 7(a)), indicating the nonsymmorphic-enforced nodal-lines may not be easily detected.

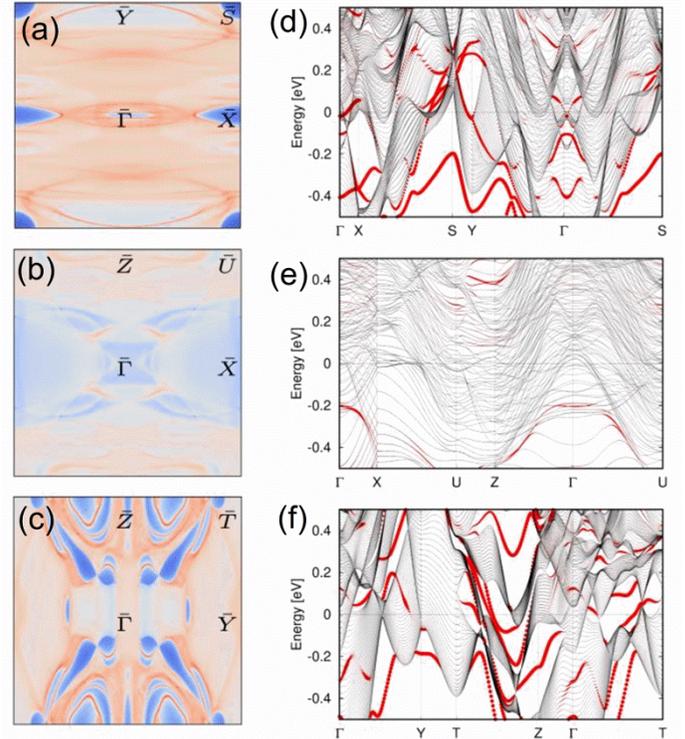

**Fig. 8.** (Color Online) Band structures of Nb$_2$P$_5$. (a)-(c) Projected surface DOSs of electronic states at the Fermi level on the (100), (010) and (110) surfaces, respectively. (d)-(f) Surface band structures of a slab with the thickness of 25 unit cell on the (100), (010) and (001) surfaces respectively along the high symmetry path of the 2D BZ. The projection of the wave function on the first layer of the slab is denoted by the thickness of the curves.

Because of the possible nontrivial topology of the inverted bands forming the nodal-line, one can carefully check the surface states at the boundary. The calculated projected spectra at the constant energy 0 for (100), (010) and (001) surfaces are presented in Figs. 8(a)-(c). For the (010) surface, the bulk DOS expands over the whole BZ, while for the (100) and (001) surfaces there are patches which lack of bulk electronic states. Especially, for the (001) surface one can find clearly surface states crossing the empty zone of projected spectrum and end at touching points of electron and hole pockets. For other two surfaces, those surface state bands are heavily entangled in the DOS of bulk bands, making their observation unlikely. To resolve this, we plotted in the Figs. 8(d)-(f) the band structures along the high symmetry paths of the 2D surface BZ of a slab system with 25 unit cell. The projected weight of the electronic states onto the surface unit cell is denoted by the thickness of the curve, from which the surface states are clearly recognized. Because of the broken inversion symmetry on the surface and also the lacking of global gap in the band structure, the surface state of nodal-line will not be as stable as those of topological insulators. It should be noted that before a careful classification, it is not conclusive about the topological nature of the surface states. To establish the exact relation between the superconductivity and the complicated electronic band structure, more work should be carried out.

## CONCLUSIONS

In conclusion, our work has demonstrated that the high-pressure technique is actually efficient in the synthesis of phosphorus rich metal phosphides. The high-pressure grown $Nb_2P_5$ crystal is a conventional weakly coupled type-II $s$-wave superconductor indicated by the superconducting parameters derived from magnetic and electrical properties measurements. Though the bands in $Nb_2P_5$ near the Fermi level are mainly denoted by the $d_{xz}$, $d_{yz}$ and $d_{x2-y2}$ orbits of Nb and the hybridization with the $p$ orbits of P seems unlikely a significant role in dominating the superconducting properties, it is still valuable to carry out further experiments to investigate the role of the peculiar crystal structure which could be viewed as the zigzag $P_\infty$ chain-inserted carrier compensated semimetal $NbP_2$. The study would be instructive for the design of more phosphorus rich metal phosphides superconductors which could provide extraordinary opportunities for discovering unconventional superconductivity. Furthermore, though the topological nature of the unveiled band crossings and electronic surface states by the theoretical calculations remain inclusive, the study on the electronic band structure of $Nb_2P_5$ would be helpful for future design of metal phosphides superconductors which could host nontrivial topological states and even topological superconductivity.

## ASSOCIATED CONTENT

### Supporting Information

Details for the high-pressure synthesis technique, the EDS data of $Nb_2P_5$ and the refinement results of the crystal structure, and the schematic layout of electrode on the sample for resistivity measurement.

### AUTHOR INFORMATION

#### Corresponding Author

*yuzhh@shanghaitech.edu.cn(ZHY),
*qfliang@usx.edu.cn(QFL),
*guoyf@shanghaitech.edu.cn (YFG).

### Author Contributions

#X.L.L. and Z.H.Y. contributed equally.

### Notes

The authors declare no competing financial interest.

### ACKNOWLEDGEMENT

The authors acknowledge the National Natural Science Foundation of China (Grant No. 11874264, 51772145), the strategic Priority Research Program of Chinese Academy of Sciences (Grant No. XDA18000000), and the Key Scientific Research Projects of Higher Institutions in Henan Province (19A140018). Y.F.G. acknowledges the starting grant of ShanghaiTech University and the Program for Professor of Special Appointment (Shanghai Eastern Scholar). J.G.Z. thanks the support by the Natural Science Foundation of Heilongjiang Province (Grant No. A2017004). The authors thank the support from Analytical Instrumentation Center (# SPST-AIC10112914), SPST, ShanghaiTech University.

TOC graphic

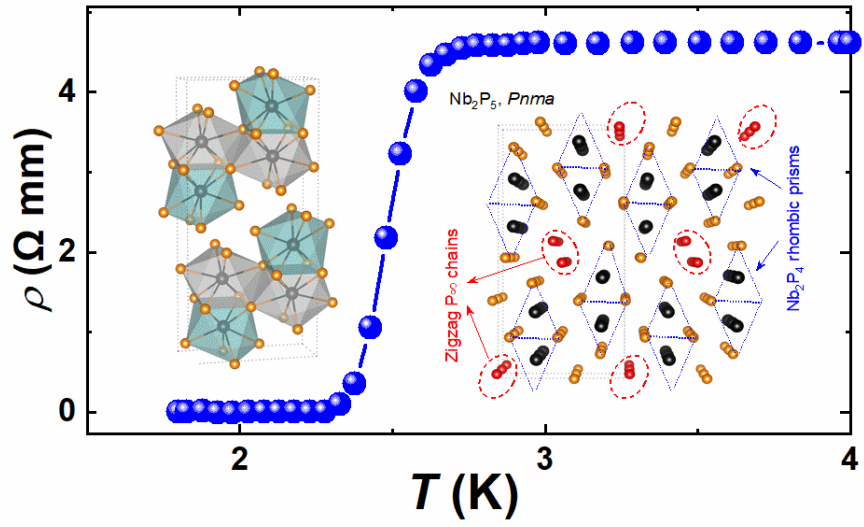

# *Supporting information*

# High-Pressure Crystal Growth, Superconducting Properties and Electronic Band Structure of $Nb_2P_5$

Xiaolei Liu[†,±,§,#], Zhenhai Yu[†,#,*], Qifeng Liang[‡,*], Chunyin Zhou[⊥], Hongyuan Wang[†], Jinggeng Zhao[l], Xia Wang[†,⊥], Na Yu[†,⊥], Zhiqiang Zou[†,⊥], and Yanfeng Guo[†,&,*]

[†]School of Physical Science and Technology, ShanghaiTech University, Shanghai 201210, China
[±]Shanghai Institute of Optics and Fine Mechanics, Chinese Academy of Sciences, Shanghai 201800, China
[§]University of Chinese Academy of Sciences, Beijing 100049, China
[‡]Department of Physics, University of Shaoxing, Shaoxing 312000, China
[⊥]Shanghai Synchrotron Radiation Facility, Shanghai Advanced Research Institute, Chinese Academy of Sciences, Shanghai 201204, China
[l]Department of Physics, Harbin Institute of Technology, Harbin 150080, China
[⊥]Analytical Instrumentation Center, School of Physical Science and Technology, ShanghaiTech University, Shanghai 201210, China
[&]CAS Center for Excellence in Superconducting Electronics (CENSE), Chinese Academy of Sciences, Shanghai 200050, P. R. China

[#]The authors contributed equally to this work.

[*]Corresponding authors:

yuzhh@shanghaitech.edu.cn,

qfliang@usx.edu.cn,

guoyf@shanghaitech.edu.cn.

1. **SHTech-2000: a multi-anvil high-temperature and high-pressure apparatus for synthesizing novel materials at ShanghaiTech University**

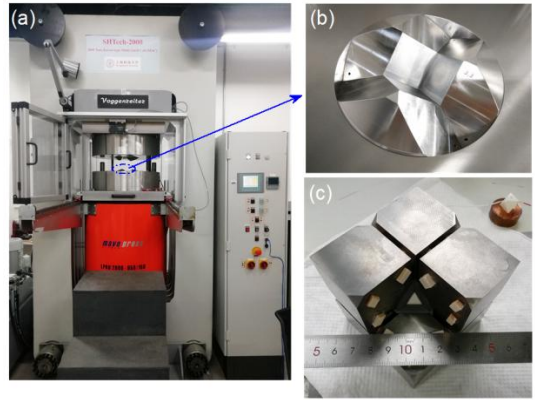

Fig. S1 (a) The optical image of Kawai-type multi anvil apparatus equipped at ShanghaiTech University (SHTech-2000). (b) The first-order stainless steel anvil. (c) The Kawai-cell composed of cubic tungsten carbide (WC) anvils, balsa soft-wood spacer, MgO octahedral pressure medium, and pyrophyllite gasket.

2. **The chemical composition analysis of sample**

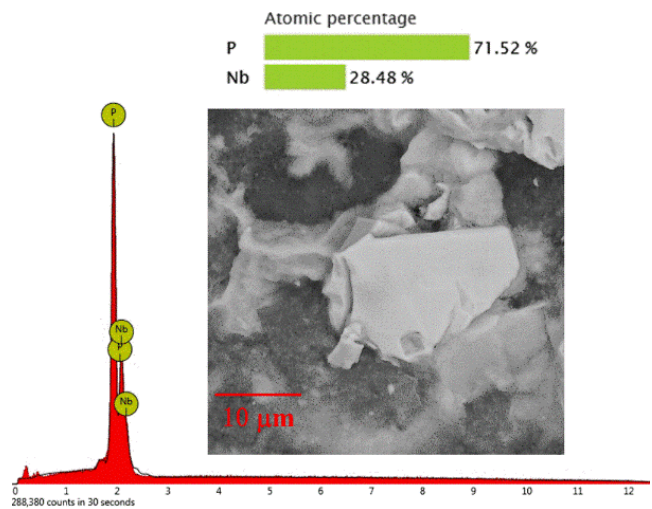

Fig. S2. The typical EDS spectrum of the $Nb_2P_5$ crystal collected at 15 kV. Insets: SEM image and atomic percentage of sample.

The atomic percentage of $Nb_2P_5$ was measured by using a Phenom ProX desktop scanning electron microscope. To get a more convincing result, we chose several different micro-crystals which are cracked from a bulk crystal for the analysis of the chemical compositions. The result shows that there are only the Nb and P elements in the crystals and the percentage of Nb is between 26% and 29%, which is very close to the stoichiometric ratio of that in $Nb_2P_5$. The insets in Fig.S2 are one typical image of the micro-crystals and its atomic percentage.

**Table S1 Crystal data and structure refinement for $Nb_2P_5$ at 150.0 K**

| | |
|---|---|
| Empirical formula | $Nb_2P_5$ |
| Formula weight | 340.67 |
| Temperature | 150.0 K |
| Crystal system | Orthorhombic |
| Space group | *Pnma* |
| Unit-cell dimens | $a = 16.7266(10)$ Å, $\alpha = 90°$ |
| | $b = 3.3453(2)$ Å, $\beta = 90°$ |
| | $c = 7.9093(6)$ Å, $\gamma = 90°$ |
| Volume Z | 442.57(5) Å$^3$, 4 |
| Density (calcd) | 5.113 g/cm$^3$ |
| Absorp coeff | 6.788 mm$^{-1}$ |
| *F* (000) | 628.0 |
| Crystal size (mm$^3$) | $0.12 \times 0.08 \times 0.05$ |
| Radiation | Mo K$\alpha$ ($\lambda = 0.71073$ Å) |
| $2\theta$ range for data collection | 4.87 ° to 63.038 ° |
| Index ranges | $-17 \leq h \leq 24$, $-3 \leq k \leq 4$, $-7 \leq l \leq 11$ |
| Reflections collected | 4034 |
| Independent reflections | 767 [$R_{int} = 0.0207$, $R_{sigma} = 0.0194$] |
| Data/restraints/parameters | 767/0/44 |
| Goodness-of-fit on $F^2$ | 1.198 |
| Final *R* indexes [$I \geq 2\sigma (I)$] | $R_1 = 0.0164$, $wR_2 = 0.0319$ |
| Final *R* indexes (all data) | $R_1 = 0.0209$, $wR_2 = 0.0330$ |
| Largest diff.peak/hole / *e* (Å$^{-3}$) | 0.64/-0.63 |

Nb$_2$P$_5$ crystals were grown by reacting Nb and red phosphorous under high-pressure conditions in a Kawai-type multi-anvil high-pressure apparatus. A selected crystal was tested on the single crystal X-ray diffractometer equipped with a Mo $K_\alpha$ radioactive source ($\lambda = 0.71073$ Å) at room temperature. Using Olex2,[1] the structure of Nb$_2$P$_5$ was solved with the SHELXT[2] structure solution program and was refined with the SHELXL[3] refinement package using Least Square minimization. Crystal refinement results were shown in Table S1.

(1)    Dolomanov, O. V.; Bourhis, L. J.; Gildea, R. J.; Howard, J. A. K.; Puschmann, H. OLEX2: A Complete Structure Solution, Refinement and Analysis Program. *J. Appl. Crystallogr.* **2009**, *42* (2), 339–341.

(2)    Sheldrick, G. M. SHELXT - Integrated Space-Group and Crystal-Structure Determination. *Acta Crystallogr. Sect. A Found. Crystallogr.* **2015**, *71* (1), 3–8.

(3)    Sheldrick, G. M. Crystal Structure Refinement with SHELXL. *Acta Crystallogr. Sect. C Struct. Chem.* **2015**, *71* (1), 3–8.

# 3. The detailed information (size and direction) of sample for electrical and magnetic transport measurement.

The sample used for electrical transport measurements is a very dense polycrystalline sample which is obtained by pressing several Nb$_2$P$_5$ crystals under the pressure of 6 GPa without heating. The samples used for resistivity and magnetic susceptibility measurements are 1.5 mm × 1 mm × 0.8 mm and 2 mm × 1.5 mm × 1 mm, respectively. The electrical current in resistivity measurement and magnetic field in susceptibility are all along the long side of the polycrystalline sample.

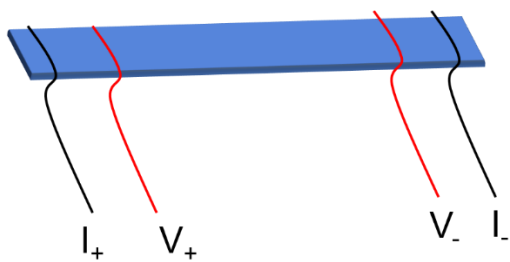

Fig. S3. The schematic layout of electrode on $Nb_2P_5$ sample in resistivity measurement.